# STATISTICAL APPROACH TO FLOW STRESS AND GENERALIZED HALL–PETCH LAW FOR EQUILIBRIUM POLYCRYSTALLINE MATERIALS

**Alexander A. Reshetnyak[1]**

Tomsk State Pedagogical University, 634041, Tomsk, Russia,
National Research Tomsk State University, 634050, Tomsk, Russia,
Institute of Strength Physics and Materials Science of SB RAS, 634055, Tomsk, Russia

**Abstract.** *A theory of flow stress, including the yield strength, $\sigma_y$ is proposed for the class of PC materials with equilibrium defect structure* (*EDS*), which is *established in the PC material after series* of *$N_0$ similar treatments of severe plastic deformation at fixed temperature T and characterized by stabilized scalar dislocation density (SDD) and average grain size d. We calculate both the stationary SDD $\rho(b,d,T)$ and suggest a way to calculate ε-evolution of an equilibrium SDD $\rho_\varepsilon$ in PC sample under quasi-static loading depending on the average size d of a grain in the range of $10^{-8}$–$10^{-2}$m, on grain boundaries orientation. The analytical dependence is realized within a disclination-dislocation mechanism in approximation of single dislocation ensemble for given phase and T. It is based on a statistical model of Boltzmann-like distribution (smoothly dependent on a strain ε) for discrete energy spectrum in each grain of a single-mode one-phase PC material with respect to quasi-stationary levels under plastic loading with the highest level equal to the energy of dislocation with maximal length. The difference of equilibrium SDD, $\rho_\varepsilon$-ρ, leads to a flow stress from the Taylor strain hardening mechanism containing (for ε=0.002) the normal and anomalous Hall–Petch relations for coarse and nanocrystalline grains, respectively, and gains a maximum at flow stress values for an extreme grain size $d_0$ of the order of $10^{-8}$–$10^{-7}$m. The maximum undergoes a shift to the region of larger grains for decreasing temperatures, revealing temperature-dimension effect. Coincidence is well established between the theoretical and experimental data on $\sigma_y$ for the materials with EDS with BCC (α-Fe), FCC (Cu, Al, Ni) and HCP (α-Ti, Zr) crystal lattices with closely packed grains at T=300K.*

## 1. Introduction

One of the main purposes of research in materials science is dedicated to controlling the internal defect substructure of grains, the structural-phase contents of polycrystalline (PC) materials in order to achieve its best strength and plasticity. An optimization of these properties is impossible without taking advantage of new technologies, the best known of which are the methods of severe plastic deformation (SPD) and their combinations with recrystallization annealing, vapor deposition, helical rolling, etc. [1]. These technologies allow for ample variations in orientation and linear size *d* of the elements of material microstructure, varying from meso-polycrystalline and coarse-grained (CG, 10–1000 μm) to fine-grained (FG, 2–10 μm), ultrafine-grained (UFG, 0.5–2 μm), sub-microcrystalline (SMC, 100–500 nm), and down to nano-crystalline (NC, <100 nm) samples. Experimental research for the mechanical properties of PC materials (micro-hardness $H$, yield strength $\sigma_y$ ultimate stress $\sigma_S$ and strain hardening coefficient $\theta$) reveals the properties of the hardening mechanism in the transition to UFG, SMC and NC states for a given material. Systematic research for the influence of the (micro)structure parameters of a certain material on the strength properties under quasi-static (QS) deformations was initiated in Refs. [2,3] by the empirical Hall–Petch (HP) relation.

$$\sigma_y(d) = \sigma_0 + kd^{-1/2} \tag{1}$$

($\sigma_0$ and $k$ are the respective frictional stress for dislocations moving inside the grains, and the HP coefficient), observed at the initial stage of the yield surface (Fig. 1c) in the diagram $\sigma$=$\sigma(\varepsilon)$ (Fig.1a), for materials either with grains of different size values (e.g. for Cu in Fig. 1b), or at the formal value $\sigma_y(d)_{|\varepsilon=0.002} \equiv \sigma_{0.2}(d)$ without a pronounced yield surface. This research was continued by many groups, e.g. in the works of R.Armstrong, H. Conrad, U.F. Kocks, G. Langford, A.W. Thompson, J.G. Sevillano, S.A. Firstov, B.A. Movchan, V.I. Trefilov, Yu.Ya. Podrezov, V.V. Rybin, V.A. Likhachev, R.Z. Valiev, V.E. Panin, E.V. Kozlov, N.A. Koneva, L.B. Zuev [4,5]. For UFG, SMC and NC samples, the relation (1) shows a significant deviation, which requires a modification [6]

$$\sigma_y(d) - \sigma_0 = k_1 d^{-1/2} + k_2 d^{-1} \tag{2}$$

by a term quadratic in $d^{-½}$ taking into account the parabolicity of the plot ($d^{-½}$, $\sigma_y(d)$), as well as a maximum at the yield strength associated with a so-called "negative value" of the HP coefficient $k$, $k$=$(\partial\sigma_y)/(\partial(d^{-½}))$ , in the region of the anomalous HP relation (see [7] and references therein). There are quite a few models intended to justify the feasibility of either the standard "linear" or the "quadratic" HP relation, on a basis of empirical approaches. Their peculiar feature [5] is a boundary hardening of grains by dislocation ensembles, including so-called triple and quadrupole joints of grains, related with their contribution to (1), (2) and also with the concept [8-10] of increased curvature-torsion of a crystal lattice (CL). In the study of PC aggregates with two-phase materials, the problem of analyzing the behavior of flow stress (FS) as a function of the size of a grain (as the main hard phase) and as an effect of grain boundaries (GBs) at the soft (second) phase becomes more involved (the contribution of the soft phase increases to tens of percent when transiting to SMC and NC materials [11]) and was examined for metal, metal-ceramic, and ceramic materials in the review [4]. Since the production of a single-mode grains of materials is technologically difficult, it leads to making allowance for distributions with respect to the grain size in a sample, and therefore takes into account the specifics of calculations for plastic and strength parameters. For such samples, beyond the relation (2) for in the case of a lognormal grain size distribution, a different model of the grain size dependence was proposed [12] for samples coated with Cr by magnetron sputtering [13]. The model takes into account a deviation from the strictly quadratic dependence (2) for $d<d_{cr2}$ by using the S-integrals technique combining three relations: (1) for $d_{cr1}<d$ ($d_{cr1}=(k_1/k_2)^2\leq$ 0.5μm, (2) for $d_{cr2}\leq d<d_{cr1}$ ($d_{cr2}\leq$0.1μm), and (3) the new relation (for $d<d_{cr2}$)

$$\sigma_y = \left(1-\left\{\frac{d-t}{d}\right\}^2\right)\sigma_{gb} + \left\{\frac{d-t}{d}\right\}^2 \sigma_{rh}, \tag{3}$$

where $t$, $\sigma_{gb}$, $\sigma_{rh}$ are the respective thickness, ultimate stress of GB, and theoretical ultimate stress of a grain. The assumption of a bi-quadratic dependence on $d^{-1/2}$, for large values ($\sigma_{gb}$, $\sigma_{rh}$)= (2;12) GPa, enables to go over to the NC region, where the anomalous (inverse) HP law holds true (see Refs. [14–18]), with increasing for $\sigma_y$ when $d$<100 nm.

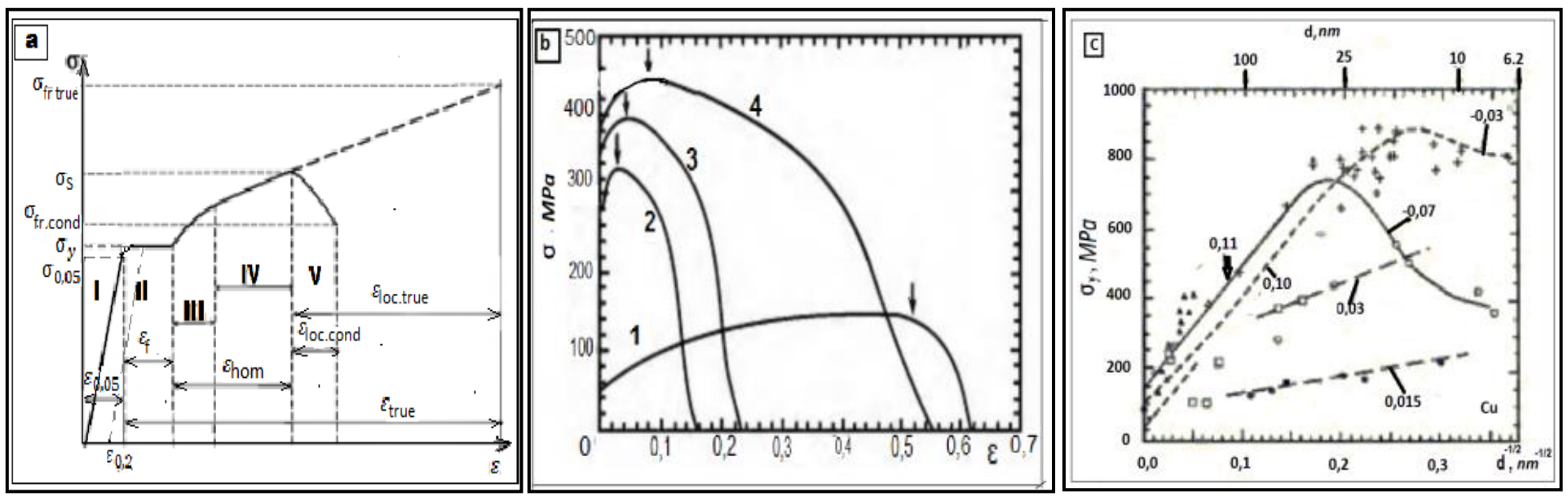

**Fig. 1** The deformation curves σ=σ(ε) for ferritic steel [5] under tensile strain in Fig. 1a, indicating: (I) quasielastic, (II) flow, (III) parabolic, and (IV) linear hardening; prefracture (V) stages with strain values and conditional limits of elasticity, yield strength, etc. Fig. 1b shows (omitting stage I) the deformation curves for Cu at T = 295 K, at the tensile rate $d\varepsilon/dt$=$10^{-2}s^{-1}$ for (1) CG, (2) after cold-rolled (60%), (3) nanostructured Cu after equal-channel angular pressing (ECAP) with 2 passes, and (4) ECAP with 16 passes [6]. The arrows indicate the maximum-ultimate stresses $\sigma_S$. Fig. 1c shows $\sigma_y$ for Cu with the normal for $k$=0.15,…,0.015 MPa·m$^{-1/2}$·and the anomalous HP laws for $k$=-0.07…,-0.03 MPa·m$^{-1/2}$ in the NC region [7].

Among the theoretical models leading to a simultaneous description of the normal and anomalous HP laws for $\sigma_y$ and $H$, one can distinguish, first of all, "*the mixed model of the plasticity of PC metals, supplementing dislocation plasticity inside the grains by the mechanism of slipping along the GBs*", based on the Maxwell strong viscous liquid within a molecular dynamics simulation for Cu and Al [19]. Secondly, the dislocation kinetic model of G.A. Malygin [20,21] on the basis of the first-order evolution equation for the average scalar dislocation density (SDD) $\rho=\rho(t)$ in the grains of the single-mode PC sample,

$$\frac{d\rho}{d\gamma} = \frac{\beta}{bd} - (k_a + k_b)\rho, \quad k_b = 4\eta_b \frac{D_{gb}}{\dot{\gamma} d^2}, \ \eta_b \approx \frac{Gb^3}{k_B T}, \tag{4}$$

following the Taylor's strain hardening mechanism [22]. In deducing (4), one supposes [20,21] that the time dependence $\rho=\rho(\gamma(t))$ is implicit through uniaxial tensile strain (or compression) $\varepsilon=\gamma/m$, $d\rho/dt=(d\gamma/dt)(d\rho/d\gamma)$ at a constant strain rate $d\varepsilon/dt=(d\gamma/dt)/m$ and shear strain rate $d\gamma/dt=b\rho u$, where $b$, $u$, $m$=3.05 are the module of the Burgers vector, dislocation velocity, Taylor orientation factor, and ($\beta$, $k_a$, $k_b$; $D_{gb}$, $G$, $k_B$, $T$) are, respectively, the coefficients determining the intensity of dislocations accumulated in a grain volume and the annihilation of screw and edge dislocations, the GB diffusion coefficient, the shear modulus, the Boltzmann constant, and absolute temperature. The model implements a competitive process of proliferation and annihilation of dislocations which depends on a quite large number of external parameters. One also considers some models with 3D dynamics of discrete dislocations [23-25] and etc.

The general conclusions drawn from the theoretical and experimental research known to date with respect to FS and $\sigma_y$ are as follows:

(1) the maximum of $\sigma_y$ is reached in some materials at certain values of the grain diameter $d_0$ in the NC region at a given $T$ and plastic deformation (PD) rate $d\varepsilon/dt$,

(2) $d_0$ is shifted to the region of larger grains with increasing values of $T$ [19] and, independently, with decreasing values of $d\varepsilon/dt$,

(3) in the regions of coarse and NC grains, there is no physical model describing simultaneously the normal and anomalous HP laws, based on a statistical approach to the spectrum of grain energies considered as the main (solid) phase of PC materials with a fixed PD, depending on the distribution of dislocation ensembles.[1]

The aim of the article is to elaborate a model for the emergence and evolution of an (quasi -) equilibrium defect structure (EDS), including 0*D*-dimensional (nanopores, bi-nanopores, etc.) and 1*D*-dimensional (dislocations) defects in the grains of a loaded PC aggregate with an EDS. To this aim we elaborate an statistical approach to the calculation of SDD on a base of evaluation of the energy spectrum of each grain of the PC sample, in view of an integral nature of FS and $\sigma_y$. *The first principal task is to analytically calculate the SDD for a single-mode PC aggregate with ESS developed in it after series of SPD treatments. The second task is to determine a evolution of SDD for a such PC sample under QS tension loading. The third and fourth tasks are to find analytical form of HP law and FS $\sigma(\varepsilon)$.*

The present study involves the introduction (in the next section) of a model to evaluate the distribution of grain energy in a PC one-phase single-mode material with an EDS with respect to quasi-discrete levels to be achieved after series of given SPD treatments for the same PC sample. To do it the approximation of single dislocation ensemble realization within all grain sizes is used. A value of SDD in a such PC material is calculated, first, after series of SPD treatments and, second, at the process of QS loading at a fixed value of accumulated strain $\varepsilon$ in assumption of homogeneous (in average) change of the EDS distribution. In the Sec. 3 we deduce the analytic form of FS $\sigma=\sigma(\varepsilon)$, as well as a generalized HP law for $\sigma_y$ and also study the corresponding asym-ptotes for CG and NC PC samples with EDS. The fulfillment of the generalized HP law for a number of metallic single-mode one-phase PC materials with a different CL is verified by means of graphic representations (Sec. 4). The article is completed by Conclusion.

## 2. Statistical Model of Grain Energy Distribution under QS PDs. Generalized Equilibrium Flow Stress and Hall–Petch Laws for Yield Strength

Let's consider a model of PC single-mode metal aggregate of volume $V$ with a shear modulus $G$, with some (BCC, or FCC, or HCP, etc) CL composed from the closely packed grains in the form of polyhedra (of diameter $d$) with some distribution of the grains through-out the sample which determine its disorientation with respect to GBs. The PC aggregate is chosen at a fixed phase state to be constant within the temperature range [$T_1$,$T_2$], see Fig.2. The smallest Burgers vector of an arbitrary (full or partial) dislocation is equal to $b$.

[1] We assume that both the stacking faults and the twin type defects (2*D* defects) prevailing in NC materials are always produced by the combination of dislocations that allow one to interpret them as combined dislocations

After multiple repeated application of the same procedure of SPD to a PC sample for given external temperature $T$ a *stable distribution of the grain's defect structure* (*EDS*) is detected. It is characterized by the stabilized (with respect to above repeated SPD procedures) average grain size $d$ and by the dislocation ensembles with Burgers vectors $\boldsymbol{b_1,\ldots,b_k}$ (see Fig. 4). Such stable defect structure should be described by a so called *equilibrium* SDD (ESDD) in the single-mode one-phase PC samples (being different from one grain to another). It is depending on $d;\ b_1,\ldots,b_k,\ T$ as $\rho=\rho(b_1,\ldots,b_k,d,T)$ and may be explicitly calculated. Note, a representation for ESDD (consistent with experiment) in the CG limit ($d>>b$) was found by H.Conrad [5,31] for single dislocation ensemble: $\rho|_{d>>b}=<\varepsilon>/(bd)$ for T=300K, with the strain degree $<\varepsilon>=\max_i \ln(c_i/d_i)$, $i=x,y,z$; for $d_i(c_i)$ the sizes of the grain before (after) SPD applied to PC sample. To determine analytically ESDD let us list the dislocation energy characteristics.

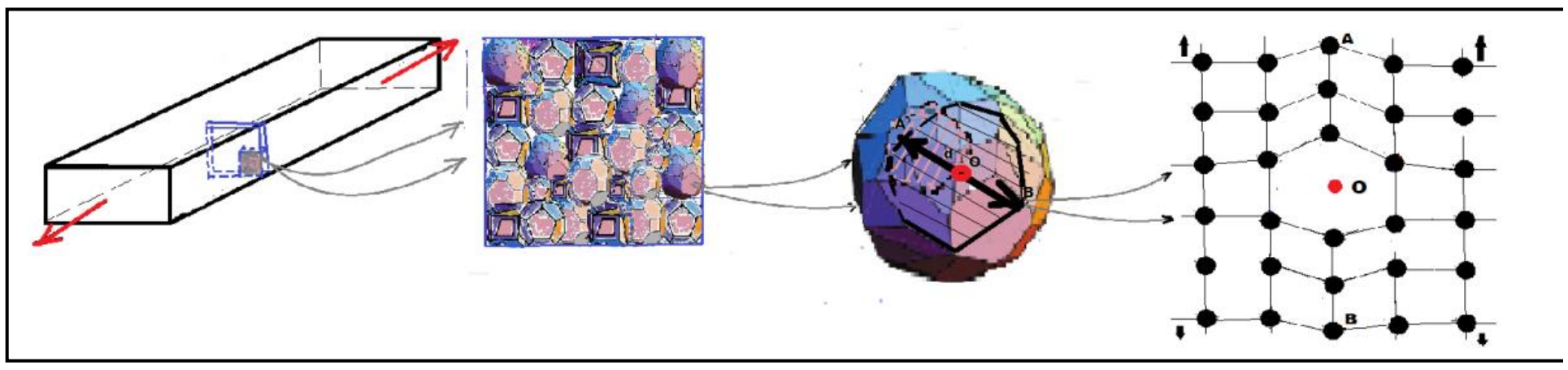

**Fig. 2** A multilevel model for a PC sample with uniformly sized (single-mode) grains and a crystallographic slip plane passing through the center of a grain.

### *2.1. Properties of dislocation energy*

Dislocations produce elastic stress fields (with a tensor $\sigma_{ik}$, $i,k$=1,2,3), which define a field of elastic strain (with a tensor $u_{ik}$) in a grain, so that the free energies $F$ of screw, edge, and mixed dislocations of length $L$ with the Burgers vector $b$ are calculated by the rule [27,28] ($F=(1/2)\int\sum_{i,k=1}^{3}u_{ik}\sigma_{ik}dV$), as follows,

$$E_d^{screw}=\frac{Gb^2L}{4\pi}ln\left(\frac{R}{r_0}+Z\right),\ E_d^{edge}=\frac{Gb^2L}{4\pi(1-\mu)}ln\left(\frac{R}{r_0}+Z\right),\quad E_d^{mix}=\frac{Gb^2L}{4\pi K}ln\left(\frac{R}{r_0}+Z\right).\quad(5)$$

Here $\mu$, $R$, $r_0$, $Z$ are, respectively, the Poisson ratio of the material (0.1<$\mu$<0.4), the radii of the dislocation zones (the cut-off parameter $R$, $R=n\cdot 10^4 b$, $n=1,2,\ldots$) and of the dislocation core (axis) $r_0\approx 3b$, the correction constant 1<$Z$<3 and $(1-\mu)\leq K\leq 1$.

The energy values of edge and screw dislocations are nearly the same and obey to:

(1) the energy of a dislocation is proportional to the dislocation length $E_d\sim L$;

(2) the energy of a *unit dislocation* of length $L_e=b$ at $R\sim L_e$ equals to $E_d^{L_e}=(1/2)Gb^3$, and for a PC material with $(G,b)$=(30GPa, $3\cdot 10^{-10}$m) it equals to $E_d^{L_e}(G,b)=2.53eV$;

(3) the energy of a dislocation which consists of $n$ elementary segments for partial dislocations, or containing $(n+1)$ atoms in the dislocation axis for full one, equals to the sum of energy values of $n$ unit dislocations $E_d^L = nE_d^{L_e}$;

(4) a dislocation with a smaller module of its Burgers vector $b_1$ of the same length $L$ as a dislocation with $b=mb_1$ ($m$>1) is more energetically preferable to emerge, since $E_d^L(G,b) = m^2 E_d^{L_e}(G,b_1)$;

(5) the energy of a unit dislocation exceeds by two orders the energy of thermal fluctuations of an atom for $T$ = 300 K: $E_d^{L_e}(G,b)/k_BT = (2.53eV)/(0.026eV)$;

(6) the energy $E_d^{L_e}$ with the smallest Burgers vector is comparable with the activation energy of an atom $E_d^{L_e} \approx E^{act}$ in the course of diffusion. Indeed, the diffusion coefficient reads $D = D_0 e^{-(E^{act}/k_BT)}$, with the frequency factor $D_0$ for most of the metals [29] being $E^{act} \in [1\text{ eV}, 4\text{ eV}]$, so for Cu, α-Fe, Nb at T=300$K$ see e.g. Table 1;

**Table 1** Comparison of the activation energy per atom and the energy of unit dislocation for Cu, α-Fe, Nb (with lattice constant $a$)

| **Material** | **α-Fe** | **Cu** | **Nb** |
|---|---|---|---|
| ***b, nm*** | $\frac{\sqrt{3}}{2}a$ =0,248 | $a/\sqrt{2}$ =0,256 | $a$=0,330 |
| ***G*, GPa** | 82,5 | 44 | 37,5 |
| $E_d^{L_e} = \frac{1}{2}Gb^3$, ***eV*** | 3,93 | 2,31 | 4,21 |
| $E^{act}$, ***eV*** | 3.05 | 2,05 | 4,13 |

(7) for a polyhedron grain inscribed in a sphere of diameter $d$, the largest rectilinear dislocation is located in one of the equatorial slip planes passing through the center of the grain, and the largest loop dislocation coincides with the equator of the polyhedron slip plane (Fig. 4), having the respective lengths L, $L_1$ and energies $E_d^N$; $E_d^{\pi N}$:

$$(L; L_l) = (N; \pi N)b,\ \left(E_d^N;\ E_d^{\pi N}\right) = (1;\pi)\frac{Gb^2 d}{4\pi K}\ln\left(\frac{R}{r_0} + Z\right) = (1;\pi)\cdot\frac{Gb^3 N}{4\pi K}\ln\left(\frac{R}{r_0} + Z\right), \quad (6)$$

where $N=[d/b;\ \pi d/b]$ is the number of atoms (segments) on the corresponding dislocation axes, and the square brackets denote the integer part of the ratios $d/b$ and $\pi d/b$.

It follows, first, that an $n$-nanopore (as uncomplete edge dislocations) and two parallel dislocations each having $n$ atoms in the axis are comparable from the energy evaluation. Second, it is advantageous to realize dislocation ensembles under PDs by crystallographic slip systems with the smallest Burgers vector $\boldsymbol{b}$, including partial dislocations, as for the materials with FCC or BCC lattices. From the item (6), it follows an approximate assumption that an energy of arbitrary dislocation may be estimated analytically using the activation energy (determined numerically or experimentally) of the atoms that form the axis.

*2.2. Derivation of equilibrium scalar dislocation density.*

Let us call the single-mode one-phase PC sample by ***equilibrium PC sample*** (or by one with ***EDS*** if in it after series from $N_0=N_0(\alpha;T)$ identical combinations (denoted by the letter $\alpha$) of mechanical tests at SPD for the temperature T for all tests with finite integers $N$: $N \geq N_0$, it is established the defect structure with $\rho=\rho(b_1,...,b_k,d,T)$: $|\rho(N)-\rho(N_0)| << \rho(N_0)$ in the grains and the average grain's size $d(N)$ is stabilized: $|d(N)- d(N_0)| << d(N_0)$.

**Remark:** In the real mechanical tests at SPD the constant value of temperature T is conserved in the volume of the sample only in average, also the *single-mode* property is realized only locally in the PC sample (as for torsion on the Bridgemen's anvils). Such separate equilibrium parts of the PS sample can be mechanically extracted.

Let's deduce the stationary ESDD $\rho=\rho(b_1,...,b_k,d,T)$ in such equilibrium PC sample in approximation of single dislocation ensemble (for least $b_1=b$) realization for all sizes $d$.

The following points, that determine the statistical model for description of the EDS energy levels distribution induced by dislocation mechanism, are crucial, describing mathematically a probability space $(\Omega, U.P)$ of events for the grains of given PC sample:

(1) The spectrum of mechanical energy for each grain consists of discrete levels, $E_d^0, E_d^1, E_d^2, \cdots, E_d^n, \cdots, E_d^N$, ..., starting with the lowest energy level $E_d^0$ of an ideal crystallite, followed by the level $E_d^1$, with a unit dislocation, and then by the levels $E_d^2,\ldots,E_d^n$, with respective dislocations (axes) of length $2b,\ldots nb$, and finally by the level $E_d^N$, being the energy of the maximal dislocation (6). For each elementary PD act, the grain either acquires or loses (local restoration of crystallinity) a $1D$-defect with $n$ atoms in its axis and the energy values $E_d^n$ at $n = 0,1,\ldots,N,\ldots$. (For each PD act, it is possible to expect the appearance of (curvilinear) dislocation with a large number $\widehat{N}>N$ of atoms in the axis);

(2) Each grain may be in a state with defect structure characterizing by the $N$-valued vector $(m_1, m_{2,\ldots,} m_n, \ldots, m_N)$, with $m_1$ unit dislocations, $m_2$ dislocations having 3 atoms in their axis (consisting of 2 segments),..., $m_n$ dislocations having $(n+1)$ atoms in their axis (consisting of $n$ segments),..., $m_N$ maximal rectilinear dislocations with the mechanical energy $\sum_{n=1}^{N} m_n E_d^n$ of all dislocations and no allowance made for the energy of elastic deformation;

(3) the probability for any of possible defects origin within SPD series to occur as we examine the state of EDS of a grain for given $T$ within approximation of the equidistant energy spectrum with a step to be equal to the energy of a unit dislocation in accordance with the Boltzmann distribution (for $\varepsilon$=0):

$$\Delta E_{n+1,n}=E_d^{n+1} - E_d^n = (1/2)Gb^3,\ n=0,1,\ldots,N=[d/b] \quad (7)$$

$$P(E_n) \equiv P_n(0) = \mathrm{A}(0)e^{-\frac{\frac{1}{2}Gb^3}{k_B T}\frac{E_n}{E_N}} = \mathrm{A}(0)e^{-\frac{\frac{n}{2}Gb^3}{Nk_B T}}, \mathrm{A}(0) = \frac{e^x-1}{e^x},\ x = \frac{\frac{1}{2}Gb^3}{k_B T}\frac{b}{d} \quad (8)$$

From the $(\Omega, U.P)$ structure for the model of PC sample in question it follows the values for the average energy $\langle E_d \rangle$ of a dislocation and the number $\langle n_d \rangle$ of atoms in its axis (segments), following the rule of ensemble averaging, according to (8),

$$\langle E_d \rangle = \mathrm{A}(0)\sum_{n=0}^{N} E_n e^{-M\frac{E_n}{E_N}} = \tfrac{1}{2}Gb^3\left(e^{Mb/d}-1\right)^{-1},\ M = \tfrac{1}{2}Gb^3/k_B T, \quad (9)$$

$$\langle n_d \rangle = \mathrm{A}(0)\sum_{n=0}^{N} n e^{-M\frac{E_n}{E_N}} = \left(e^{Mb/d}-1\right)^{-1} \equiv f_N(b,d,T),, \quad (10)$$

where the factor $(2\pi K)^{-1}\ln(R/r_0+Z)$ in $E_n$=$(n/2)Gb^3$ has been omitted. The quantity $\langle n_d\rangle$ coincides with the probability distribution function $f_N$ $(b,d,T)$ for the occurrence of a dislocation of energy $\langle E_d\rangle$ in an arbitrary grain of the PC sample with EDS for fixed $T$. In the limit for CG, SMC and NC materials with a finite value $N$~$10^2$–$10^3$ and also for grain diameters $d\leq 5nm$, the respective estimations for the energy (9) are as follows

$$\left\{\lim_{d\gg b}, \lim_{d\twoheadrightarrow bM}, \lim_{d/b\ll M}\right\}\langle E_d\rangle = \frac{1}{2}\left\{\frac{2d}{b}k_BT = \frac{N}{M}Gb^3, Gb^3\left(e^{Mb/d}-1\right)^{-1}, Gb^3e^{-Mb/d}\right\}. \quad (11)$$

It implies that the first value is equal to the thermal energy of $N=d/b$ atoms, being the energy of a dislocation with $NM^{-1}(\sim 10^{-2}N)$ atoms on its axis, i.e., in average a dislocation is adjacent to the GB from inside; the second value: $E_d^{Le} < \langle E_d\rangle < 10E_d^{Le}$ describes the fact of complete "germination" of dislocations in grains of SMC and NC materials; and the third value $\langle E_d\rangle_{|(M/ln2)>\mathrm{N}} < 0.5Gb^3$ for $(M/ln2)b >$ d, due to $\mathrm{f}_N < 1$ corresponds to the absence (in average) in such grains of dislocation emergence and also of 0$D$ defects, which leads to the unhardening of an equilibrium PC sample. The latter means that within SPD process the collective thermal energy of atoms in a grain for given $T$ is insufficient to generate a dislocation.

The average dislocation length $<L_\mathrm{d}>=b\mathrm{f}_N(b,d,T)$ and the sum of the lengths of all dislocations in an arbitrary grain (in average) with an *average dislocation number* $N_e$:

$$L_\Sigma = N_e <L_\mathrm{d}> = N_e\, b\, \mathrm{f}_N(b,d,T), \quad (12)$$

determine the stationary ESDD $\rho=\rho(b,d,T)$ under SPD at the crystalline phase (far from GB) of the sample with allowance made for approximation of any grain as a 3D-sphere,

$$\rho = L_\Sigma/\mathrm{V_d} = N_e\, 6\mathrm{b}/(\pi d^3)\left(\mathrm{e}^{\mathrm{Mb/d}}-1\right)^{-1}. \quad (13)$$

The value of $N_e$ should depend on the energy scale $M$ and a number of segments in dislocation, $d/b$, for the PC sample with an EDS. Thus, in an arbitrary grain we find $N_e$ from the limiting case of $\rho(b,d,T)$ in CG aggregates according to H.Conrad [5,31]:

$$\lim_{\mathrm{d}>>\mathrm{b}}\rho(b,d,T)=<\varepsilon>/(bd) \;\; => \;\; N_e=<\varepsilon>M(\pi d)/(6\mathrm{b})\,. \quad (14)$$

As the result, $\rho = <\varepsilon>M/(d^2)\left(e^{Mb/d}-1\right)^{-1}$. We stress, that ESDD, in fact, does not depend on the specific orientation of the GBs. Whereas, the FS (in particular $\sigma_y$) essentially depends on GB angles distribution in a PC sample. To take into account this factor we introduce a model of so-called quasi-equilibrium PC sample, when calculating the SDD. To do so, we divide the sample on parts (not, in general, connected) with subvolumes $V_i$, $i$=1,2,...,$k$, $\sum_j V_i=V$, where the average angle $0\leq\alpha_i<\pi/2$ of GB disorientation is restricted within small conus ($\alpha_i$, $\alpha_i+\Delta\alpha_i$) subject to the ordering, $\alpha_i<\alpha_{i+1}$. Such division may be determined by means of EBSD (also by SEM, energy dispersive X-ray) analysis (see, Fig. 3) applied to the real PC sample with respect to grain size and GB angle distributions [40-43].

Thus, the DS of the quasi-equilibrium PC sample induced by only the most probable dislocation ensemble should be determined by the ESDD $\rho=\rho(b,d,T)$ and distribution of GB disorientation ($\alpha_i$, $N_i$), $i=1,...,k$ for a weight factor $N_i= V_i/V$ ($\sum_I N_i=1$).

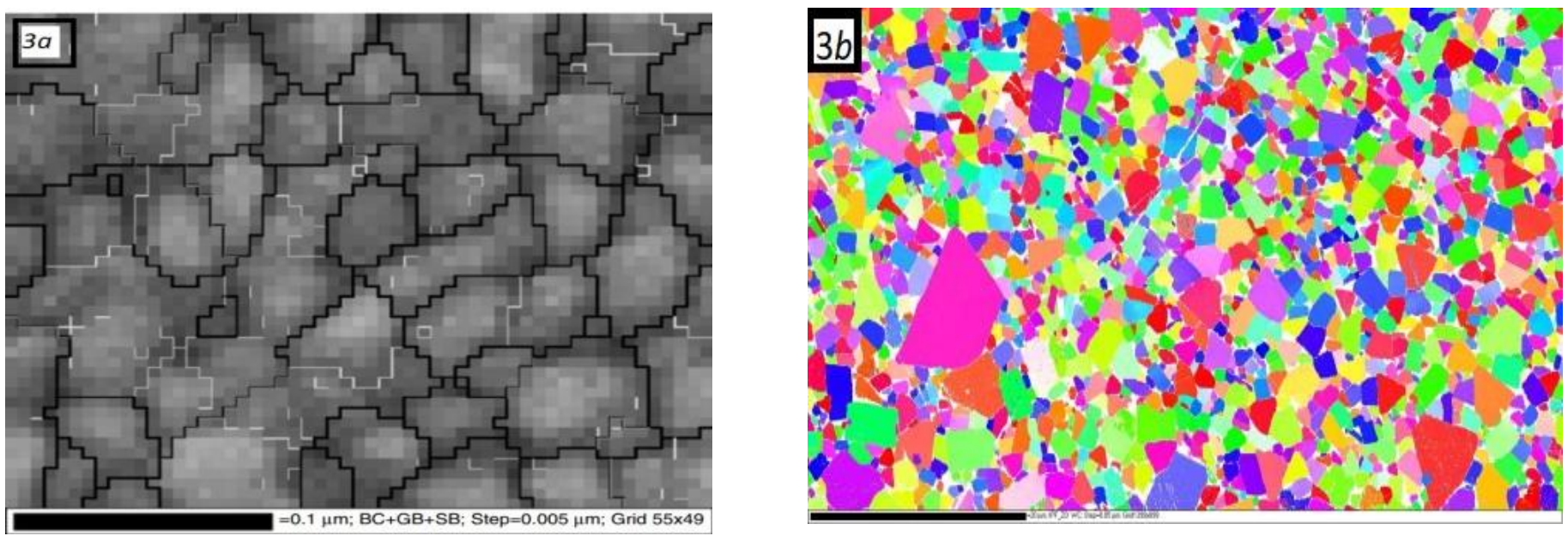

**Fig. 3** EBSD map of GB angle distribution in EBSD of Pt thin film (Fig. 3*a*) with high-angle (black) and low-angle CBs (white) in almost single-mode PC sample with grain size d~30-40nm (from [46]). ESBD orientation map of WC/Co (Fig. 3*b*) (from [52]) for grain size for 0.5 μm <d<20μm and G distributions .

*2.3. Evolution of ESDD under Quasi-static loading*

To provide the evaluation of SDD within QS loading (by tension), as $\rho_\varepsilon=\rho(b,d,T;\,\varepsilon)$ we enlarge our assumptions (1)-(3) to include *t*- (*equally ε-*) *dependence*

(4) We suppose, that the distribution of the probabilities $P_n(\varepsilon)$, $n$=0,1,2,... for an origin of the respective defects at an *elementary PD act* in a grain to occur at the time instant $t=(\varepsilon-\varepsilon_{0.05})/\dot{\varepsilon}$ as we examine the EDS of a sample (beyond the conditional elastic limit $\varepsilon_{0.05}$) corresponding to the approximation of a equidistant grain spectrum of item (1) for ($\Omega, U.P$) satisfies to the linear homogeneous equations

$$\frac{d}{d\varepsilon}P(E_n,\varepsilon)=f_n(\varepsilon)P(E_n,0)\Leftrightarrow\frac{d}{dt}P(E_n,\dot{\varepsilon}t)=\dot{\varepsilon}f_n(\dot{\varepsilon}t)P(E_n,0). \quad (15)$$

The solution of (15) $P(E_n,\varepsilon)=F_n(\varepsilon)P(E_n,0)$, for $\ln F_n(\varepsilon)=\int_0^\varepsilon f_n(y)dy$ ($F_n=1$ if $f_n=0$), following to an idea of self-similarity for the EDS evolution under QS loading, thus preserving the Boltzmann-like distribution for any $\varepsilon>\varepsilon_{0.05}$, may be chosen with $\tilde{f}_n(\varepsilon)=\frac{d}{d\varepsilon}[A(\varepsilon)A^{-1}(0)e^{-\frac{nGb^3}{2Nk_BT}((1+\varepsilon)^3-1)}]$, as

$$P(E_n,\varepsilon)\equiv P_n(\varepsilon)=A(\varepsilon)e^{-\frac{\frac{1}{2}Gb_\varepsilon^3}{k_BT}\frac{E_n}{E_N}}=A(\varepsilon)e^{-\frac{\frac{n}{2}Gb_\varepsilon^3}{Nk_BT}},\quad A(\varepsilon)=\frac{e^x-1}{e^x},\quad x=\frac{\frac{1}{2}Gb_\varepsilon^3}{k_BT}\frac{b}{d}. \quad (16)$$

Here, we introduce the *effective Burgers vector* $b_\varepsilon$, $b_\varepsilon=b(1+\varepsilon)$, and also the *effective dislocation energy* $E_n(\varepsilon)=(n/2)G(b_\varepsilon)^3$ where account is taken of $E_n(0)=E_n$ and $E_d^n(\varepsilon)/E_d^N(\varepsilon)\equiv E_n(\varepsilon)/E_N(\varepsilon)=E_n(0)/E_N(0)$.

To evaluate a number of dislocation $N_0(\varepsilon)$ in a grain at the moment $t=\varepsilon/\dot{\varepsilon}$ we introduce

(5) *a minimal time* $\Delta t_0$ between PD acts under which a grain increases its PD from the value $\varepsilon$ to $\varepsilon+\Delta\varepsilon$, for stretching along the *z*-axis and thereby enlarged by the value $d(1+\varepsilon)\Delta\varepsilon=B\cdot b$ (for $B=1$),[2] is connected with the movement of an dislocation or with (*elementary PD act* itself) the appearance of screw dislocation or of two edge dislocations (with the *Burgers vectors* ***b*** and **-** ***b***) situated in the crystallographic slip plane passing, e.g. according to energy of average dislocation (9) near GB of a grain. For the time instant $t = (\varepsilon - \varepsilon_{0.05})/\dot{\varepsilon}$ the interval $\Delta t_0$ is determined by the condition,

$$\varepsilon - \varepsilon_{0.05} + \Delta\varepsilon = \dot{\varepsilon}(t + \Delta t_0) \Rightarrow \Delta t_0 = \Delta\varepsilon/\dot{\varepsilon} = b/[\dot{\varepsilon}\, d(1+\varepsilon)] \quad (17)$$

[2] In the initial paper [30] we have considered a case $d(1+\varepsilon)\Delta\varepsilon=b_\varepsilon$ therefore with $B=(1+\varepsilon)$ in the equation (17).

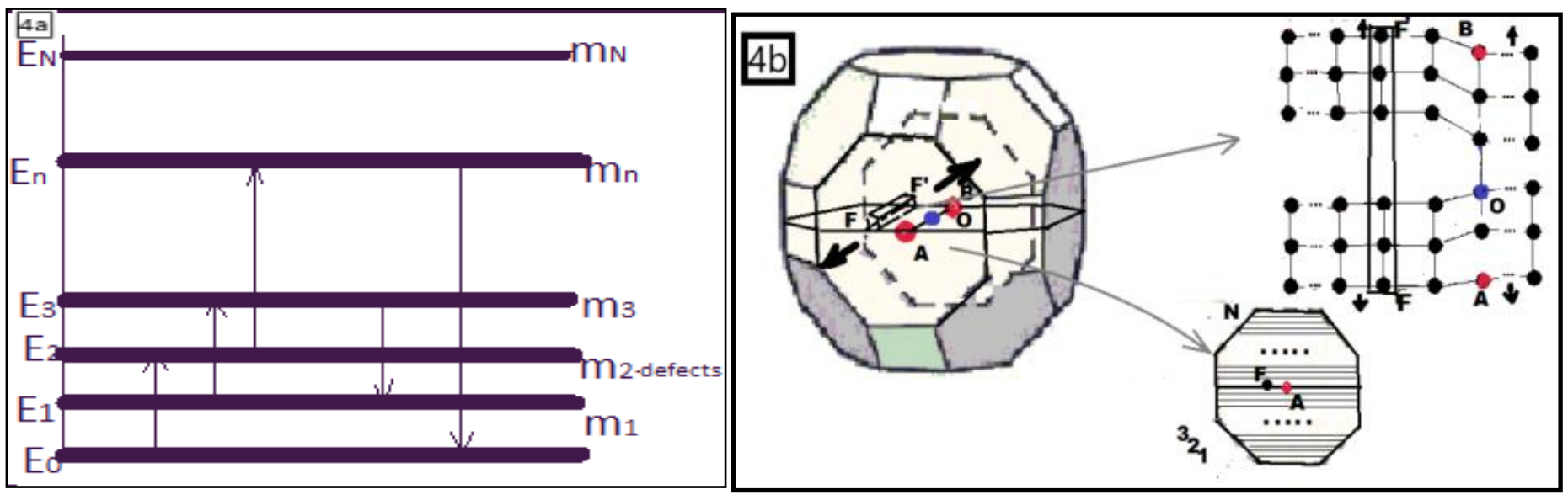

**Fig. 4** An equidistant grain energy spectrum drawn (due to thermal atomic oscillations) by thick lines far removed from the GB in Fig. 4a (energy gaps near the boundary are narrowed),. An arrow pointed from $E_m$ to $E_n, E_n > E_m$ ($E_m > E_n$) at a given instant $t$ indicates a transition with an increase (decrease) in the grain energy as the DS is modified according to (7). A variant for specification of the *polyhedral parameter* $m_0 = m_0(N)$ (shown on Fig. 4b) with a number $N$ of identical parallel slip planes coincident with the direction of loading (short black arrows) and separated from each other by $b$. The crystallographic plane *FAF'* contains an axis of maximal straight (rectilinearr) dislocation *FA*, which coincides with the (thickened) central line.

Deformations with the value $\Delta\varepsilon$ occur in almost all of the crystallographic planes of the grain spaced by the distance $nb$, $n$=1,2,…,[$d/b$]. The minimal number of dislocations $N_e(\varepsilon)$ being in an arbitrary grain [arising both after SPD at $\varepsilon$=0: $N_e(0)=N_e$ and then within QS loading for $\varepsilon$>0: $N_e(\varepsilon)=N_e+N_0(\varepsilon)$] during the time interval $t$ in order to achieve the residual PD value $\varepsilon$ is determined by the relation for *ε-dependent part of* $N_e(\varepsilon)$ (taking into account edge dislocation emergence in pairs):

$$N_0(\varepsilon) = (k_e+k_s)\, t/\Delta t_0 = (k_e+k_s)\, \varepsilon(1+\varepsilon)\, d/b = m_0\, [M(\pi d)/(6b)]\, \varepsilon(1+\varepsilon)\, d/b\, . \quad (18)$$

Here, first, we naturally suppose on homogeneous character of the form of dependence for both terms $N_e$ and $N_0(\varepsilon)$ with account for CG limit (14). Second, the quantities $k_e$ ($k_e=2p$), $k_s$ are respectively the coefficients for accounting of edge and screw dislocations and $m_0$ is a dimensionless *polyhedral parameter* [30], to be determined from comparison with SDD (or yield strength) for concrete PC sample in CG limit and may be interpreted as the number of planes contributing to the deformation of a grain (Fig. 4b).

Again, repeating the steps (9)-(13) when deducing stationary $\rho=\rho(b,d,T)$ we obtain total ESDD $\rho_\varepsilon=\rho(b,d,T;\varepsilon) = L_\Sigma(\varepsilon)/V_d(\varepsilon)$ in the form:

$$\rho_\varepsilon = N_e 6b/(\pi d^3)\left(e^{Mb/d} - 1\right)^{-1} + N_0(\varepsilon)6b/(\pi d^3)\left(e^{M(\varepsilon)b/d} - 1\right)^{-1} + o(\varepsilon^2)$$

$$= \langle\varepsilon\rangle(M/d^2)\left(e^{Mb/d} - 1\right)^{-1} + m_0\, \varepsilon\, (M/d^2)\left(e^{M(\varepsilon)b/d} - 1\right)^{-1} + o(\varepsilon^2) \quad (19)$$

with allowance for a change in the grain volume $V_d(\varepsilon)=(1/6)\pi d^3(1+\varepsilon)(1-1/2\varepsilon)^2=V_d(0)+o(\varepsilon)$ under PD. Note, that ε-independent part of $\rho(b,d,T;\varepsilon)$ coincides with stationary ESDD (13). The EDS of single-mode PC sample under QS loading is determined by $\rho=\rho(b,d,T;\varepsilon)$ and distribution of ε-dependent GB disorientation ($\alpha_i(\varepsilon)$, $N_i$), $i=1,...,k$. In the limits of CG and NC aggregates for small PDs, the values of $\Delta_\varepsilon\rho=\rho_\varepsilon-\rho$ are evaluated as

$$\left(\lim_{d\gg b}, \lim_{d\twoheadrightarrow bM(\varepsilon)}\right)\Delta_\varepsilon\rho=\frac{m_0}{bd}\varepsilon\left((1+\varepsilon)^{-2},\ (Mb/d)\left(e^{M(\varepsilon)b/d}-1\right)^{-1}\right), \qquad (20)$$

which holds true for experimentally observed ESDDs (for $m_0$~$10^1$–$10^2$) [5,31].

Let's consider the isotropic grain's distribution in the PC sample, in question, e.g. with a cubic CL. It implies that the distribution of crystallographic slip planes inside the angle $[-\pi/4,\pi/4]$ relative to the loading alone $z$-axis is such that the average number of dislocations $<N_0(\varepsilon)>$ for ε-dependent ESDD in an average grain equals to $N_0(\varepsilon)/\sqrt{2}$. The case of anisotropic distributions of grains (as well as with textures) makes it necessary, first, to introduce both a *texture factor* $K=K(x,y,z)$, when calculating the average number $\overline{N}_0(\varepsilon)$

$$\overline{N}_0(\varepsilon)=\langle K\,N_0(\varepsilon)\,\rangle_V,\quad 0<K\leq 1$$

over all grain configurations in a PC sample (or, equally GB disorientation ($\alpha_i(\varepsilon)$, $N_i$)) and, second, to take account for the external angle among the directions of QS loading with respective integral direction of CL for all grains.

**Remarks**:

1) To determine $m_0$ within the model approximation one should to express it from experimentally defined values of ESDD for single-mode PC samples with closely packed grains, or in CG limit as $m_0^{CG}$ or in NC limit as $m_0^{NC}$ according to the rule (20), e.g. for ε=0.002. Thereby, by one from the criteria of the validity of the statistical approach for determination of ESDD would be checking of: $m_0^{CG}=m_0^{NC}$;

2) In both CG and NC limits (20) we have used an assumption on absence of GB disorientation, i.e. for $\alpha_i(\varepsilon)\approx 0$;

3) Parameter $m_0$ may be independently determined from generalized HP law for yield strength, see below Eqs. (23), (24), in the CG and NC limits as $\overline{m}_0^{CG}$ and $\overline{m}_0^{NC}$ and then compared as $\overline{m}_0^{CG}=\overline{m}_0^{NC}$ (as well as it provides the check of Taylor hardening law validity: $(\overline{m}_0^{CG},\overline{m}_0^{NC})=(m_0^{CG},m_0^{NC})$);

4) In case of real PC materials an approximation of single dislocation ensemble realization (equidistant spectrum of grain's energy) within all grain's sizes is modified by considering of new dislocation ensembles, which correspond to origin of twining (2D defect) structure. In this case the dislocations arise with different admissible Burgers vectors for a given CL, so that the grain energy spectrum should consist of discrete bands, $E_d^0, E_d^{1k}, E_d^{2k},\ldots, E_d^{nk}\ldots, E_d^{Nk}$, parameterized by the number 1, …, $n$ of segments in the dislocation axes, and by the number $k$ of different vectors $b_1<b_2<\ldots<b_k$. Instead of probabilities (8) one may consider its multi-ensemble analog

$$P\left(E_{n_1},\ldots,E_{n_k}\right)=\mathrm{A}(0)\prod_{j=1}^{k}\mathrm{e}^{-n_j M_j\frac{b_j}{d}},\ \mathrm{A}(0)=\frac{\mathrm{e}^{y}-1}{\mathrm{e}^{y}},\ y=\sum_{j=1}^{k}M_j\frac{b_j}{d},\ M_j=\frac{\frac{1}{2}\mathrm{G}b_j^3}{k_B\mathrm{T}} \qquad (21)$$

5) The distributions (13), (19) can be traced from an quasi-particle interpretation of grain energy changing [30].

## 3. GENERALIZED EQUILIBRIUM FLOW STRESS AND HALL–PETCH LAW

We suppose that Taylor deformation (dislocation) hardening law [22],which holds for CG materials due to the interaction energy of dislocations for tangential FS, $\tau\sim Gbl^{-1}\sim Gb\rho^{-1/2}$, is also valid both for UFG and NC materials (following Ref. 20) and also for $\varepsilon > 0{,}002$

$$\tau(\varepsilon)=\tau_f+\alpha Gb(\Delta_\varepsilon\rho)^{-1/2}, \qquad \Delta_\varepsilon\rho=\rho_\varepsilon-\rho. \tag{22}$$

Here $\alpha$ and $\tau_f$ are respectively a dislocation interaction constant $\alpha$, varying for different materials in the range (0.1–0.4) and frictional stress at the interaction of propagating dislocations with lattice defects and obstacles of non-deformation origin. Taking into account that the FS of a PC sample, $\sigma(\varepsilon)$, is proportional to $\tau(\varepsilon)$, $\sigma(\varepsilon)=m\tau(\varepsilon)$, according to (19) and (22), we obtain for isotropic GB distribution

$$\sigma(\varepsilon)=\sigma_0(\varepsilon)+\alpha mG\frac{b}{d}\sqrt{\frac{m_0}{\sqrt{2}}\varepsilon\frac{Gb^3}{2k_BT}}\,\left(e^{M(\varepsilon)b/d}-1\right)^{-\frac{1}{2}}+o(\varepsilon),\ \sigma_0(\varepsilon)=m\tau_f(\varepsilon) \tag{23}$$

Expression (23) provides – with accuracy up to higher orders in $\varepsilon$ – the main analytical result of applying our statistical model to the determination of FS at the crystalline phase of a single-mode equilibrium PC sample with isotropic GB orientation in all grain ranges under QS loading (within single dislocation ensemble approximation)[3].

For general form of GB disorientation ($\alpha_i(\varepsilon)$, $N_i$), the integral FS $\sigma_\Sigma=\sigma(\varepsilon,\{\alpha_i\})$ for PC sample with quasi-EDS may be determined additively according to

$$\begin{aligned}\sigma_\Sigma&=\sigma_0(\varepsilon,\{\alpha_i\})+\textstyle\sum_i N_iK(\alpha_i)\alpha mG\frac{b}{d}\sqrt{m_0\varepsilon_iM(0)}\,\left(e^{M(\varepsilon_i)b/d}-1\right)^{-\frac{1}{2}}+o(\varepsilon)\\&=\sigma_0+\textstyle\sum_i N_iK(\alpha_i)\sigma_i(\varepsilon_i)+o(\varepsilon_i),\ \text{for}\quad \varepsilon=\ \sum_i N_i\varepsilon_i,\quad \sum_i N_i=1\,,\end{aligned} \tag{24}$$

where nonnegative numbers $K(\alpha_i)$ ($K(\alpha_i)\le 1$) determine the value of GB disorientation (from EBSD analysis) in respective $N_i$ part of PC sample with strength $\sigma_i$ and strain $\varepsilon_i$. We suppose the results (23), (24) are applicable for quasi-equilibrium PC samples not only for yield strength but as well for the stages of parabolic and linear hardening down to the stages of pre-fracture according to the Backofen-Considére condition. Notice, that the values $\sigma_0(0)=\sigma(0)=0$ and $\sigma_0(0.002)=\sigma_0$ are given for $\varepsilon$=0.002 in (1), (23), (24). Dependence $\sigma(\varepsilon)$ (23), determines the FS maximum $\sigma_m(\varepsilon)=\sigma(\varepsilon)|_{d=d_0}$ as depending on the extremal grain size $d_0$ based on a transcendental equation implied by $\partial\sigma/\partial d=0$,

$$\partial\,\sigma(\varepsilon)/\partial\,d=\frac{Q(\varepsilon)}{d^2}(e^x-1)^{-\frac{3}{2}}\left[e^x-1-\frac{1}{2}xe^x\right]=0,\ \text{for}\ x=M(\varepsilon)b/d \tag{25}$$

for a certain $Q(\varepsilon)$ independent of $d$. With accuracy up to five digits the solution for equation (25) is $x$ =1.59363, and therefore

$$d_0(\varepsilon,T)=b\frac{Gb^3(1+\varepsilon)^3}{2\cdot1.59363\cdot k_BT}\,. \tag{26}$$

The FS maximum $\sigma_m(\varepsilon)$ for a PC aggregate without a second (soft) phase is calculated as

$$\sigma_m(\varepsilon) = \sigma_0(\varepsilon) + \alpha m G\sqrt{\frac{m_0}{\sqrt{2}}\varepsilon\frac{d_0(\varepsilon,T)\cdot 1.59363}{b(1+\varepsilon)^3}}\left(e^{1.59363}-1\right)^{-\frac{1}{2}} = \sigma_0(\varepsilon) + Kd_0^{1/2} \quad (27)$$

with inverse dependence of $\sigma_m$ on $d_0$ as compared to HP relation (1) for any $\varepsilon$.

---

[3] In case of multiple dislocation ensemble due to probabilities (21) the $\varepsilon$-dependent part of ESDD $\Delta_\varepsilon\rho=\sum_j\Delta_\varepsilon\rho_j$ (and therefore FS) should have more complicated form $\Delta_\varepsilon\rho\sim m_0\sum_j\varepsilon_j(M_j/d^2)\left(e^{M_j(\varepsilon_j)b_j/d}-1\right)^{-1}$ and presents a separate problem.

For CG materials, the normal HP law for FS when $\varepsilon$=0.002 implies a relation between HP coefficient $k(\varepsilon)_{|\varepsilon=0.002}$ and $\bar{m}_0^{CG}$ in the CG limit for the polyhedral parameter $m_0$:

$$\sigma(\varepsilon)_{|d\gg b} = \sigma_0(\varepsilon) + k(\varepsilon)d^{-\frac{1}{2}},\quad k(0.002) = \alpha(\bar{m}_0^{CG})^{\frac{1}{2}} mG\sqrt{\frac{\varepsilon}{\sqrt{2}}b\frac{M(0)}{M(\varepsilon)}}_{|\varepsilon=0.002}$$
$$\Longrightarrow\ \alpha^2\bar{m}_0^{CG} = \frac{\sqrt{2}}{\varepsilon b}\frac{k^2(\varepsilon)}{(mG)^2}\frac{M(\varepsilon)}{M(0)}_{|\varepsilon=0.002} \quad\Longleftrightarrow\quad \bar{m}_0^{CG} = \frac{\sqrt{2}}{\varepsilon b}\frac{k^2(\varepsilon)}{(\alpha mG)^2}\frac{M(\varepsilon)}{M(0)}_{|\varepsilon=0.002} \quad (28)$$

This correspondence (with assumption $\bar{m}_0^{CG} = \bar{m}_0^{NC} = m_0$ from Remark 3) permits to establish explicit interrelation among the theoretical and empirical HP relations for number of materials and to reveal a *temperature–dimension effect* (see as well [30]).

## 4. Generalized Hall–Petch law for equilibrium α-Fe, Cu, Al, Ni, α-Ti, Zr.

To implement theoretical HP law in the graphic form, we use the experimental data of Table 2 on the HP coefficient $k(0{,}002)$ applied for single-mode PC samples with BCC, FCC and HCP CL, with the values of $\sigma_0$, $G$, $a$ [33], Burgers vector with the least possible length $b$ for the respective most probable dislocation sliding systems (Table 3), the dislocation interaction constant $\alpha$ [20,33] the calculated values of the least dislocation $E_d^{L_e}$, extreme size $d_0$, maximal difference of $\sigma_y$: $\Delta\sigma_m$ according to (23) and (28) for $T$=300K

**Table 2**: The values $\sigma_0$, $\Delta\sigma_m$=($\sigma_m$-$\sigma_0$), $E_d^{L_e}$, $k$, $m_0$, $\alpha$ in BCC, FCC and HCP polycrystalline metal samples with $d_0$, $b$, $G$ obtained using (26) and [46] at ε=0.002.

| CL | BCC | FCC | | | HCP | |
|---|---|---|---|---|---|---|
| **Material** | **α-Fe** | **Cu** | **Al** | **Ni** | **α-Ti** | **Zr** |
| ***σ*₀, *MPa*** | 170 (annealed) | 70 (anneal.); 380 (cold-worked) | 22 (anneal. 99,95%); 30(99,5%) | 80 (annealed | 100(~100 %);300 (99,6%) | 80-115 |
| ***b, nm*** | $\frac{\sqrt{3}}{2}a$=0.248 | $a/\sqrt{2}$=0.256 | $a/\sqrt{2}$ =0.286 | $a/\sqrt{2}$ =0.249 | $a$=0.295 | $a$=0.323 |
| ***G, GPa*** | 82.5 | 44 | 26.5 | 76 | 41.4 | 34 |
| ***T, K*** | 300 | 300 | 300 | 300 | 300 | 300 |
| ***k,MPa·m$^{1/2}$*** | 0.55-0.65 ($10^{-5}-10^{-3}m$) | 0.25 ($10^{-4}-10^{-3}m$) | 0.15 ($10^{-4}-10^{-3}m$) | 0.28 ($10^{-5}-10^{-3}m$) | 0.38-0.43 ($10^{-5}-10^{-3}m$ | 0.26 ($10^{-5}-10^{-3}m$) |
| ***α*** | – | 0.38 | – | 0.35 | 0.97 | – |
| $E_d^{L_e}=\frac{1}{2}Gb^3$ **, eV** | 3.93 | 2.31 | 1.96 | 3.72 | 3.33 | 3.57 |
| $\alpha^2\bar{m}_0^{CG}$ | 14.9-19.5 | 9.8 | 8.71 | 4.24 | 22.3-28.5 | 14.1 |
| ***d*₀, *nm*** | 23.6 | 14.4 | 13.6 | 22.6 | 23.8 | 28.0 |
| **Δ*σ*m, *GPa*** | 2.29-2.56 | 1.34 | 0.83 | 1. 20 | 1.58-1.70 | 1.00 |

The values for $k$ at $\varepsilon$=0.002 are used for α-Fe, Cu, Ni [5], for Zr [33], for Al [32], and for α-Ti [34,35] with the range of grain size shown in the brackets:

**Table 3**. Most probable crystallographic sliding systems at $T$=300K [32] for α-Fe, Cu, Al, Ni, α-Ti, Zr for BCC, FCC (Miller indices), HCP lattices (Miller–Bravais indices).

| | α-Fe | Cu | Al | Ni | α-Ti | Zr |
|---|---|---|---|---|---|---|
| **Plane** | {110},{112},{123} | {111} | {100},{111} | {111} | $(10\bar{1}0)$,$(10\bar{1}1)$ | $(10\bar{1}0)$ |
| **Direction** | <111> | <110> | <110> | <110> | $[2\bar{1}\bar{1}0]$ | $[11\bar{2}0]$ |

The graphic dependence $\sigma_y = \sigma_y(d^{-1/2})$ for the hard (crystalline) phase of single-mode PC aggregates of α-Fe, Cu, Al, Ni, α-Ti, Zr with closely-packed randomly oriented grains at T=300K, is shown on the Fig. 5 on the basis of the Tables 2, 3.

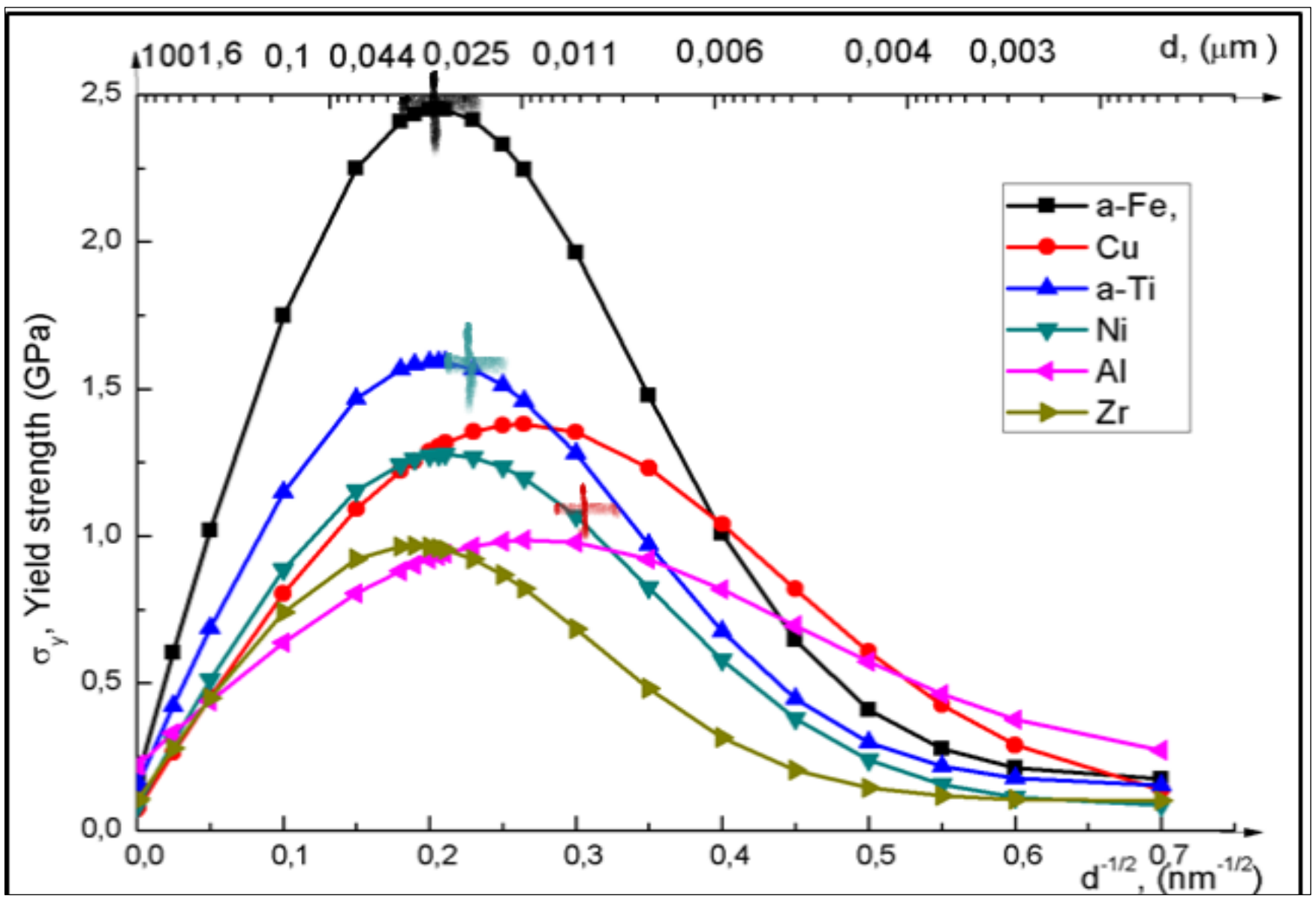

**Fig. 5.** Graphic dependence for generalized HP law with an additional upper scale with size of grains $d$ given in μm. Upper axis $d$ is changing within range (∞;0) with the inverse quadratic scale and the correspondence (100; 1,6; 0,1; 0,044; 0.025; 0,011; 0,006; 0,004; 0,003) μm ↔ (0,005; 0,015; 0,1; 0,15; 0,2; 0,3; 0,41; 0,5; 0,57) $nm^{-1/2}$ for the respective values on the lower axis.

The least possible values of the parameter $m_0(k)$ for the α-Fe,α-Ti, values of $\sigma_0$ for annealed materials with the maxima of $\sigma_y$ calculated for the extreme grain size values $d_0$ relative to Table 2. According to Fig. 5, experimental data coincide approximately at the extreme size values [5], as well as the values for maximums $\sigma_m$ [37]. However, the maximum values $\sigma_m$ requires taking into account a negative input from the GB phase, multi-mode character for experimental PC samples, as well as theirs non-equilibrium DS and other dislocation ensembles excitation.

Remarkably, the values of $d_0(0.002, T)$ for α-Fe, Cu, Al, Ni, α-Ti, Zr in the Table 2, are in complete agreement with the range (both empirical and theoretical) of critical size values for the average diameters of grains $d_{cr}$ for above PC samples (listed, e.g., in Ref. 5 [Table 2.6, pp. 110–111] and in Ref. 36), ranging from 5–10 nm to 20–50 nm, particularly, for $d_{cr}$(Cu)=10 nm≈ $d_0$(300;Cu)=14.4 nm from Ref. 18 and $d_{cr}$(Ni)=20 nm ≈ $d_0$(300;Ni)=22.6 nm. For the corresponding values for the maxima of experimental $\sigma_y$, i.e. $\sigma_m(0.002)$ in single-mode (on average) PC samples (see, e.g., Refs. 37,38), we find that $\sigma_m(0.002)$(α-Fe) ≈ 2.6±0,1 GPa, $\sigma_m(0.002)$(Ni) ≈ 1.6±0,1GPa, $\sigma_m(0.002)$ (Cu) ≈1.0±0,1 GPa coincide approximately with the theoretical maxima, with allowance for the various definitions of HP coefficients according to the literature (see Table 2 in Ref. 36 with $k$(Cu) ∈ [0.01,0.024] for UFG PC samples).

## 5. Discussion.

For a joint analytical description of the normal and anomalous HP laws for yield strength in the *PC* (metallic) *aggregates with quasi equilibrium defect structure* (*EDS*) it is proposed the model of the single-mode one-phase PC aggregate with EDS and it is developed the statistical approach to deduce the stationary ESDD $\rho(b_1,...,b_k,d,T)$ established in the PC sample after series of the same SPD tests.

The model includes the following natural assumptions:

1) EDS for given $T$ is characterized by single dislocation ensemble with the least Burgers vector $b$=$b_1$, which corresponds to the equidistant discrete energy spectrum $E^0_d$, $E^1_d$,…,$E^n_d$,…, $E^N_d$ of each grain for any value of average diameter $d$ within [$10^{-8}$-$10^{-2}$] m;

2) Boltzmann distribution (8): {exp$(-nM/N\}$, $n$=1,2,… for origin of respective dislocation (of the length $nb$) with energy scale $M=\frac{1}{2}Gb^3/k_BT$ and $N=d/b$;

3) isotropic orientation of GB with closely packed grains.

As the result, the energy $\langle E_d\rangle$ (9) of an average dislocation in arbitrary grain of PC sample in question correctly describes the behavior of the set of dislocations for CG, NC and subNC samples (11) in accordance with thermo-fluctuation mechanism within SPD. The equilibrium SDD $\rho(b,d,T)$ (13), (14) satisfies by construction to the Conrad expression (14) in the CG limit.

For the quasi-static loading (by tension) the $t$-(*equally* $\varepsilon$-) evolution of ESDD $\rho_\varepsilon=\rho(b,d,T;\varepsilon)$ is naturally based on the above assumptions for the model of PC sample enlarged by the system of *ε-dependence condition* (15), providing a homogeneous changing for probabilities $\{P(E_n,\varepsilon)\}$ with the partial solutions (16) $\{\exp(-nM(\varepsilon)/N\}$ for effective energy scale $M(\varepsilon)=\frac{1}{2}Gb^3_\varepsilon/k_BT$ and Burgers vector $b_\varepsilon=b(1+\varepsilon)$.

Note, the stationary ESDD is measurable and can be explored to determine yield strength $\sigma_y$, whereas the pure $\varepsilon$-dependent part of ESDD $\Delta_\varepsilon\rho$ in (19) was derived under evaluation of a number of dislocation $N_0(\varepsilon)$ in an arbitrary grain with help of the notion of *a minimal time* $\Delta t_0$ (17) of PD for changing of a grain (and then PC sample) along the loading direction on a quantity proportional to a module of the least Burgers vector $b$. It has permitted to derive the flow stress $\sigma(\varepsilon)=\sigma(b,d,T;\varepsilon)$ (23) for equilibrium PC sample under assumption of Taylor deformation (dislocation) hardening law validity in all grain size regions. The generalization for the integral FS $\sigma_\Sigma$ in case of general form of GB disorientation ($\alpha_i$(ε), $N_i$), i=1,…,k with angles $\alpha_i$(ε) is given by (24) in approximation of additive input of each FS $\sigma_i(\varepsilon_i)$ with the weight $N_i$.

The expressions for the yield strength $\sigma_y(d) = \sigma(0.002)$ and $\sigma_{\Sigma y}(d) = \sigma_\Sigma(0.002)$ from the FS $\sigma(\varepsilon)$ (23) and $\sigma_\Sigma(\varepsilon)$ (24) present the basic results of the statistical approach application. We stress, first, that as compared to $\sigma_y(d)$, [for whose determination according to (22), (23) one can also use the stationary ESDD (13)] the derivation of FS $\sigma(\varepsilon)$ (23) and $\sigma_\Sigma(\varepsilon)$ (24) depends on a choice of ε-dependent form for probabilities $\{P(E_n,\varepsilon)\}$, i.e. of the factors $f_n(\varepsilon)$ (15) and requires further study. Second, the generalized HP relation $\sigma_y(d)$ and FS $\sigma(\varepsilon)$ depend on the polyhedral parameter $m_0$, (see remark 3)), which may be independently determined both from generalized HP law in the CG (with help of HP coefficient (28)), NC limits as $\bar{m}_0^{CG}$, $\bar{m}_0^{NC}$ and from asymptotic behavior of ESDD in the same limits as $m_0^{CG}, m_0^{NC}$ then compared as $\bar{m}_0^{CG} = \bar{m}_0^{NC}$, $m_0^{CG} = m_0^{NC}$ (as well as it provides the check of Taylor hardening law application $(\bar{m}_0^{CG}, \bar{m}_0^{NC}) = (m_0^{CG}, m_0^{NC})$). Third, the sets of the experimental data, e.g. $\sigma_0$, HP coefficient $k$, yield strength $\sigma_y$ for the same single-mode PC samples of number $P$ for given T, but with different defect structure characterized by respective GB disorientation $(\alpha^s_{i^s}(0.002), N^s_{i^s})$, $s=1,..,P$, $i^s=1,\dots,k^s$, and by possible (nano)-porous part from the weak GB phase presence are differed. Thus, to determine parameter $m_0$ in the framework of the model (23) for pure crystalline phase in CG limit by the rule (28) one should to ordered the average angles $\langle\alpha^s\rangle$ for each PC sample as $\langle\alpha^1\rangle \leq \langle\alpha^2\rangle \leq \dots \leq \langle\alpha^P\rangle$ and to match the values of parameter $\bar{m}_{01}^{CG} \geq \bar{m}_{02}^{CG} \geq \dots \geq \bar{m}_{0P}^{CG}$.

Obtained dependence permitted to reveal the nature of the normal and anomalous HP laws for $\sigma_y$, to determine within the model approximation the extremal grain size $d_0$ (26) and maximal value $\sigma_m$= max $\sigma_y$=$\sigma_y(d_0)$ (27) for equilibrium single-mode PC sample. The value $d_0(\varepsilon,T)$ decreases in temperatures: $d_0(\varepsilon,T_1) > d_0(\varepsilon,T_2)$ for $T_1 < T_2$, thus determining in part a *temperature-dimension effect* [30,35] in PC materials. Note, that the latter property is opposite to one marked by item (2) obtained within molecular dynamics simulations [19]. In addition, $d_0(\varepsilon,T)$ provides increasing in strain: $\frac{d_0(\varepsilon_1,T)}{d_0(\varepsilon_2,T)} = \left[\frac{1+\varepsilon_1}{1+\varepsilon_2}\right]^m \geq 1$ for $\varepsilon_1 > \varepsilon_2$, which is the basis for experimental finding of $m$ to be equal to, $m = 3, 2, 1$ or 0, thus revealing $\varepsilon$-dependence for $P(E_n,\varepsilon)$, (i.e. for $M_m(\varepsilon) = (1+\varepsilon)^m \frac{1}{2} Gb^3/k_B T$).

The graphical representations (Fig. 5) of the established HP law for single-mode one-phase PC aggregates with BCC (α-Fe), FCC (Cu, Al, Ni) and HCP (α-Ti, Zr) crystal lattices with closely packed grains at T=300 K show a sufficient close coincidence between the theoretical and experimental data for both the extremal grains $d_0(0.002{,}300)$ and the maximum $\sigma_m(0.002)$ for these materials. (with some specific for the latter in view of not accounting for peculiar features of BCC. FCC, HCP CL for equilibrium PC aggregates, explicit GB regions with weak phase, considered in Refs. 30. 39, excitation at SPD of other dislocation ensembles, twinners).

The proposed approach permits the basis for generalization to the case of multi-mode (one-phase) equilibrium PC samples with closely packed grains, arbitrary GB disorientation, weak GB (with porous) part, and aspects of dispersion. We stress, that the experimental verification of the temperature-dimension effect, which includes $T$, $\varepsilon$-behavior of the extremal grain $d_0(\varepsilon,T)$ and $\sigma_y, \sigma_m$ for PC sample with EDS are very necessary.

**Acknowledgements** The author is grateful to I.L.Buchbinder, Yu.Sharkeev,V.G. Bagrov,S.G.Psakhie,E.Shilko for support and discussions. The work has been supported under the Program of fundamental research at the State Academy of Sciences for 2013-2020.